\documentclass[reprint,amsmath,amssymb,aps,prl]{revtex4-1}
\usepackage{graphicx}
\usepackage{dcolumn}
\usepackage{bm}
\usepackage{lmodern}
\usepackage{mathtools}
\usepackage{dsfont}
\usepackage{mathrsfs}
\usepackage{amsthm,amsfonts,mathtools}
\usepackage{mathbbol}
\usepackage{hyperref}
\usepackage{xcolor}

\newcommand{\rrho}{\hat{\boldsymbol{\rho}}}

\newcommand{\diag}{\mathrm{diag}}
\newcommand{\Mat}{\mathrm{Mat}}

\newcommand{\reals}{\mathds{R}}
\newcommand{\complex}{\mathds{C}}
\theoremstyle{plain}

\theoremstyle{definition}

\begin{document}

\preprint{APS/123-QED}

\title{Generalized quantum-classical correspondence for random walks on graphs}  

\author{Massimo Frigerio}
\email{Electronic address: massimo.frigerio@unimi.it}
\affiliation{Quantum Technology Lab $\&$ Applied Quantum Mechanics Group, Dipartimento di Fisica ``Aldo Pontremoli'', Universit\`a degli Studi di Milano, I-20133 Milano, Italy}
\affiliation{INFN, Sezione di Milano, I-20133 Milano, Italy}

\author{Claudia Benedetti}
\email{Electronic address: claudia.benedetti@unimi.it}
\affiliation{Quantum Technology Lab $\&$ Applied Quantum Mechanics Group, Dipartimento di Fisica ``Aldo Pontremoli'', Universit\`a degli Studi di Milano, I-20133 Milano, Italy}
\affiliation{INFN, Sezione di Milano, I-20133 Milano, Italy}

\author{Stefano Olivares}
\email{Electronic address: stefano.olivares@fisica.unimi.it}
\affiliation{Quantum Technology Lab $\&$ Applied Quantum Mechanics Group, Dipartimento di Fisica ``Aldo Pontremoli'',
Universit\`a degli 
Studi di Milano, I-20133 Milano, Italy}
\affiliation{INFN, Sezione di Milano, I-20133 Milano, Italy}

\author{Matteo G. A. Paris}
\email{Electronic address: matteo.paris@fisica.unimi.it}
\affiliation{Quantum Technology Lab $\&$ Applied Quantum Mechanics Group, Dipartimento di Fisica  ``Aldo Pontremoli'',
Universit\`a degli 
Studi di Milano, I-20133 Milano, Italy}
\affiliation{INFN, Sezione di Milano, I-20133 Milano, Italy}

\date{\today} 

\begin{abstract}
We introduce a minimal set of physically motivated postulates that the Hamiltonian $\mathrm{H}$ of a continuous-time quantum walk should satisfy in order to properly represent the quantum counterpart of the classical random walk on a given graph. We found that these conditions are satisfied by infinitely many quantum Hamiltonians, which provide novel degrees of freedom for quantum enhanced protocols, In particular, the on-site energies, i.e. the diagonal elements of  $\mathrm{H}$, and the phases of the off-diagonal elements are unconstrained on the quantum side. The diagonal elements represent a potential energy landscape for the quantum walk, and may be controlled by the interaction with a classical scalar field, whereas, for regular lattices in generic dimension, the off-diagonal phases of $\mathrm{H}$ may be tuned by the interaction with a classical \emph{gauge} field residing on the edges, e.g., the electro-magnetic  vector potential for a charged walker.
\end{abstract}
\maketitle
%
%
Continuous-time quantum walks (CTQW) on graphs are traditionally defined 
as the quantum analogue of classical random walks (RW), by promoting the classical transfer matrix, i.e. the RW graph Laplacian, to an Hamiltonian \cite{farhi1998quantum,childs2002example,mulken2011continuous}. However, this association does not encompass all the possible \emph{quantum} evolutions of a walker on a graph, thus strongly limiting the set of exploitable quantum Hamiltonians, with possible negative implications on the estimation of a quantum advantage of CTQWs vs. RWs in specific tasks\cite{ambainis2003quantum,venegas2008quantum,childs09comp,portugal2013quantum,wang2013physical,
shenvi2003quantum,childs2004spatial,chakraborty2016spatial,wong2016laplacian, abal2010spatial,foulger2015quantum}. 
A question thus arises on whether it is possible to define more general quantum walks on a graph, by considering all Hermitian Hamiltonians compatible with a given graph topology, and how do these generalized QWs compare with their classical analogues. Recently,  {\em chiral quantum walk} has been introduced \cite{cqw1,cqw4,cqw2,cqw3}, showing that complex phases in the Hamiltonian generator of a CTQW may be exploited to introduce a directional bias in the dynamics. However, without further justification, the introduction of chiral CTQWs seems to be a departure from the original spirit of quantum walks. In particular, no clear and general connection with the classical RW has emerged for chiral CTQW so far, and no interpretation of the new degrees of freedom entailed by Hermitian Hamiltonians for chiral CTQW has been discussed. 
\par
In this Letter, we shall first put in a rigorous framework the concept of continuous-time classical random walk on undirected simple graphs, and show that the associated classical probabilities cannot arise from a unitary quantum evolution via the Born rule. Having excluded this simplest equivalence, we proceed by listing a number of reasonable minimal requests (of topological, algebraic and probabilistic nature) 
for a correspondence between a RW and a CTQW on the same graph. We show that these assumptions lead to a single equation that relates the classical generator $\mathrm{L}$ of bi-stochastic transformations (the graph Laplacian) to the Hermitian quantum generator $\mathrm{H}$. As a special case, we recover the standard association $\mathrm{H} = \mathrm{L}$. However, such an equation admits infinitely many quantum Hamiltonians as solutions for a given legit classical generator $\mathrm{L}$, consistently with the common intuition that any map from classical to quantum evolutions should be one-to-many. In particular, the on-site average energies and the complex phases of the 
off-diagonal elements are unconstrained on the quantum side. Therefore, any classical RW on a graph corresponds to infinitely many \emph{chiral} CTQWs whose on-site energies are also arbitrary (an aspect that has been overlooked so far).

{We also provide a physical motivation for these additional degrees of 
freedom to appear: the diagonal elements can always be interpreted 
as a potential energy landscape for the CTQW, in other words as the interaction with a classical scalar field,  whereas the off-diagonal phases of $\mathrm{H}$ arise 
from an interaction of the CTQW with a classical \emph{gauge} field residing 
on the edges (for regular lattices in generic dimension). We are thus able to reinterpret chiral CTQW on lattices as the  Schr\"{o}dinger equation of the 
spatially discretized version for a non-relativistic particle with 
minimal coupling to a vector gauge field.} 
\par
Let us start by defining the most general continuous-time quantum walk on a finite graph. We consider a finite-dimensional Hilbert space $\mathcal{H} \simeq \complex^{n}$ with a preferred basis $ \{ \vert j \rangle \}_{j=1,...,n}$ formed by the localized states on the $n$ distinct vertices of an \emph{undirected, connected simple graph}. A Hermitian operator $\hat{H}$ acting on $\mathcal{H}$ is the Hamiltonian of a CTQW if and only if it respects the topology of the graph, that is to say if for $j \neq k$ we have 
$ \left[ \mathrm{H} \right]_{kj}  =   \langle k \vert \hat{H} \vert j \rangle  \neq  0 $
if and only if the vertices $j$ and $k$ are connected by an edge, where we introduced the matrix $\mathrm{H}$ that describes $\hat{H}$ on the localized basis. Clearly, very little about the structure of $\hat{H}$ is actually specified by the graph itself. If the starting state is a pure, localized state $\rrho_{0} =  \vert j \rangle \langle j \vert $, it will stay pure under the unitary evolution induced by $\hat{H}$:
\begin{equation}
\label{eqQev1}
    \rrho (t) \ \ = \ \ e^{-i \hat{H} t} \vert j \rangle \langle j \vert e^{ i \hat{H} t } \ \ = \ \ \vert \psi_{j} (t) \rangle \langle \psi_{j} (t) \vert
\end{equation}
where $\vert \psi_{j}(t) \rangle  =  \sum_{k=1}^{n} \alpha_{kj}(t) \vert k \rangle$ and
\begin{align}
\label{eqQev2}
    \alpha_{kj}(t) \ \ &= \ \ \langle k \vert e^{-i\hat{H} t} \vert j \rangle \ \ = \ \ \left[  e^{-i \mathrm{H} t} \right]_{kj}.
\end{align}

We would like to find a classical counterpart to the dynamics described above. In particular, we are looking for a \emph{classical} RW on the same graph. This is described by a continuous-time Markov chain, whose evolution satisfies a semi-group structure. In formulae, we seek a continuous family of stochastic $n \times n$ matrices, $\mathrm{P}(t)$ with $t \geq 0$ acting on the vector $\underline{ \mathrm{p} } \in \reals^{n}$ of probabilities associated with the sites of the graph, such that:
\begin{equation}
\label{eqCLev}
    \mathrm{P}(t_{2}) \mathrm{P}(t_{1} ) \ \ = \ \ \mathrm{P}(t_{1} + t_{2} ) \ \ \ , \ \ \ \forall t_{1}, t_{2} \geq 0.
\end{equation}
Then there exists an $n \times n$ \emph{real} matrix $\mathrm{L}$ that generates this evolution through:
\begin{equation}
\label{eqCLgen}
    \mathrm{P}(t) \ \ = \ \ e^{- t \mathrm{L}} \ \ \ , \ \ \forall t \geq 0.
\end{equation}
Again the graph topology will be imposed by the non-zero off-diagonal elements of the generator $\mathrm{L}$. In Eq.(\ref{eqCLgen}), the request that $\mathrm{P}(t)$ is left-stochastic $\forall t \geq 0$, which is necessary for a valid evolution of probabilities, implies that the sum of the entries in each column of $\mathrm{L}$ is $0$. Moreover, it is natural to pick $\mathrm{L}$ symmetric (thus $\mathrm{P}$ is bi-stochastic) to maintain a clear relation with the \emph{undirected, simple} graph. Finally, one can check that positivity of $\mathrm{P}$ is satisfied $\forall t \geq 0$ if we impose that all the off-diagonal elements of $\mathrm{L}$ are negative, while the diagonal elements have to be positive to compensate for the vanishing sum of the columns (or rows). All this construction will be assumed as a \emph{definition of a classical RW}. In the special instance of \emph{unweighted graphs}, $\mathrm{L}$ is uniquely specified by the topology and it is the \emph{Laplacian} matrix of the graph, i.e. $\mathrm{L} = \mathrm{D} - \mathrm{A}$ where $\mathrm{A}$ is the adjancency matrix and $\mathrm{D}$ is the diagonal matrix encoding the connectivities of each vertex.  Going back to quantum, the unitary evolution of the CTQW induces an evolution of the probabilities at each vertex of the graph according to Born's rule:  
$ \pi_{kj} (t) \ \ = \ \ \vert \alpha_{kj} (t) \vert^{2}$,
where $\pi_{kj}(t)$ stands for the probability of finding the particle on site $k$ at time $t$ if it is localized at site $j$ at time $0$, and it should \emph{\textbf{not}} be interpreted, a priori, as a standard transition matrix in the sense of Eq.(\ref{eqCLev}). Indeed, while it is true that if the initial state $\rrho_{0}$ is diagonal, $\rrho_{0} = \sum_{j} \rho^{0}_{j} \vert j \rangle \langle j \vert$, then $\underline{ \mathrm{p} } $ evolves according to $ \mathrm{p}_{k} (t)  =  \sum_{j} \pi_{kj}(t) \rho^{0}_{j} $, this is no longer true at intermediate times, nor if the initial state is coherent in the localized basis. In fact, we shall now show that \emph{the semigroup structure of Eq.(\ref{eqCLev}) can never be fulfilled by the quantum probabilities computed according to the Born rule}. Indeed, we are looking for a real $n\times n$ matrix $\mathrm{L}$ fulfilling the equality:
\begin{equation}
    \left[ e^{- t \mathrm{L}} \right]_{kj}  \ \  = \ \ \Big\vert \left[e^{-i \mathrm{H} t } \right]_{kj} \Big\vert^{2} \ \ \ \ \ \forall t \geq 0. \\
\end{equation}
Expanding both sides to first order in $t$, we deduce:
\[   -\  t \ \mathrm{L}_{kj}    
\ = \ -i t \ (\mathrm{H}_{jj} -\mathrm{H}^{*}_{jj} ) \delta_{jk}. \]
Since $\mathrm{H}$ is Hermitian, $\mathrm{H}_{jj} - \mathrm{H}^{*}_{jj} = 0$ and the equation above can be satisfied if and only if $\mathrm{L} = {\boldsymbol 0}$ as a matrix. 

On the other hand, we do know a way to associate a RW to a CTQW: just take $\mathrm{L} = \mathrm{H}$ and forget about the Born rule. Classical probabilities will 
evolve according to Eq.(\ref{eqCLev}-\ref{eqCLgen}), while the quantum dynamics is independently specified by Eq.(\ref{eqQev1}-\ref{eqQev2}). 
\par
However, there are a few objections to this simpler approach. Firstly, imposing $\mathrm{L} = \mathrm{H}$ forces $\mathrm{H}$ to be real, not a natural request for an Hamiltonian: this points at the fact that a classical-to-quantum correspondence should be one-to-many, therefore some further structure is required to specify \emph{all} the \emph{Hermitian} Hamiltonians associated with the given classical generator $\mathrm{L}$. Secondly, it is a rather arbitrary and artificial way to define the quantum-classical correspondence, with no physical motivation, and many other correspondences could exist (although we have just excluded that based on the Born rule). 
\par
We shall address these issues and find a generalized quantum-classical correspondence for continuous-time random walks. We start by stating reasonable requests to be fulfilled by any equation linking a Laplacian matrix $\mathrm{L} \in \Mat (n, \reals)$ and a generic Hermitian Hamiltonian $\mathrm{H} \in \Mat (n, \complex)$ that describe a RW and a CTQW, respectively, on the same undirected, simple graph with $n$ vertices (T denotes topological requests, A is for algebraic and P for probabilistic):
{\bf T0} Both $\mathrm{H}$ and $\mathrm{L}$ should preserve the topology of the underlying physical system, therefore the equation must enforce that $\mathrm{L}_{jk} \neq 0$ iff  $\mathrm{H}_{jk}
\neq 0$ ($j \neq k$); {\bf A1} Given $\mathrm{H}$, the equation must admit a unique solution for $\mathrm{L}$ with
$\sum_j \mathrm{L}_{kj}=0$  $\forall k$, so that it is a valid Laplacian matrix of a (possibly weighted) graph;
{\bf A2} The correspondence should reproduce the simple association $\mathrm{L} = 
\mathrm{H}$ whenever $\mathrm{H}$ is already a Laplacian matrix of an unweighted graph;
{\bf P3} The off-diagonal terms of $\mathrm{H}$ should be interpreted as transition amplitude rates, while those of $\mathrm{L}$ should be the corresponding transition probability rates.
{\bf P4} Given $\mathrm{H}$, the diagonals term of the solution $\mathrm{L}$ of the equation should maintain the meaning of total probabilities of leaving the corresponding sites. 

In general, for $j \neq k$, $\mathrm{H}_{jk} \in {\mathbb C}$ and
$\mathrm{L}_{jk} \in {\mathbb R}$. Then 
Condition {\bf T0} suggests:
{
\begin{equation}
\label{eqHLoffdiag}
    \mathrm{L}_{jk} \ \ = \ \ - \vert \mathrm{H}_{jk} \vert^{2}\,,
\end{equation}}
which also satisfies {\bf P3} for the off-diagonal terms. Conditions {\bf P4} and {\bf A1} for the diagonal terms may be jointly enforced by: 
\begin{equation}
\label{eqHLdiag}
    \mathrm{L}_{jj}  = 
   \sum_{s \neq j}^{n} \vert \mathrm{H}_{js} \vert^{2} =
     \langle j| \hat{H}^2|j\rangle - \langle j | \hat{H} | j \rangle ^2\,.
    \end{equation}
Condition {\bf A2} is also satisfied since one can immediately check that $ [ \mathrm{L}^{2} ]_{jj} \delta_{jk}  - \left( \mathrm{L}_{kj} \right)^{2}  =  \mathrm{L}_{kj} $
whenever $\mathrm{H}$ is a Laplacian and the simple identification $\mathrm{L} = \mathrm{H}$ is recovered.
Alternative definitions, e.g. $\mathrm{L}_{kj} = - \vert \mathrm{H}_{jk} \vert^{n}$ with $n \neq 2$, would satisfy {\bf T0} and {\bf A2}, and possibly {\bf A1}, but they would unavoidably violate {\bf P3} and {\bf P4}. Indeed, the choice $n=2$ is naturally suggested by the Born rule \emph{at the level of transition rates}: the classical rate of transition should be the square modulus of the corresponding transition amplitude. In compact form the sought equation is: 
\begin{equation}
\label{eq:LfromH}
    \left[ \mathrm{L} \right]_{kj} \ \ = \ \ [ \mathrm{H}^{2} ]_{jj} \delta_{jk}  \ - \  \mathrm{H}_{jk} \mathrm{H}_{kj}   
\end{equation}
As stated by this expression, there are many CTQWs corresponding to the same RW: the moduli of the off-diagonal entries of $\mathrm{H}$ are fixed to the square-root of the moduli of the corresponding off-diagonal elements of $\mathrm{L}$ whereas 
the phases are completely free. The diagonal elements of $\mathrm{H}$, i.e. the on-site energies of the vertices, are also unconstrained: {from a physical point of view, they do not possess a classical analogue, because the corresponding classical system is open and energy is not conserved. On the converse, by Eq.(\ref{eq:LfromH}) the diagonal elements of $\mathrm{L}$ are fixed to be the quantum fluctuations of the energy of each site}, which are also equal to the total probability of escaping each site, analogously to the connectivity for a Laplacian. {We have thus found \emph{all possible quantum walks} associated with an arbitrary classical random walk on a given graph, assuming the minimal and reasonable requests detailed above.} 
As shown in the Supplementary Material, Eq. (\ref{eq:LfromH}) may be also derived by 
adding a decoherence term with respect to the energy eigenbasis to the Von Neumann evolution equation of a chiral CTQW on a graph, thus adding physical intuition to 
the formal consistency implied by conditions {\bf T0-P4}. 
\par
The possibility of choosing \emph{complex} off-diagonal entries in the Hamiltonian $\mathrm{H}$ accounts for chirality \cite{cqw1,cqw3}, i.e. asymmetry under the time reversal transformation $t \to -t$. {We shall then keep calling \emph{chiral quantum walks} all the generalized CTQWs compatible with a given $\mathrm{L}$ through Eq.(\ref{eq:LfromH})}, even if they are more general than the original definition, since also the diagonal terms are unconstrained. Concerning classical RWs, instead, time reversal is not meaningful because $e^{-t \mathrm{L}}$ is guaranteed to be a stochastic matrix only if $t \geq 0$, and this is directly related to the irreversible dynamics of classical RWs. We deduce that, whenever $\mathrm{H}$ is real, the quantum-classical comparison is unambiguous under time reversal: there is just one possible choice of time direction for the classical walk, and both choices for the quantum walk are equivalent. In contrast, when $\mathrm{H}$ is complex the quantities $\vert \langle k \vert e^{-i \mathrm{H} t} \vert j \rangle \vert^{2}$ are \emph{not} in general symmetric under $t \to -t$, and a possible ambiguity in the quantum-classical comparison arises. This is resolved when considering all possible Hamiltonians $\mathrm{H}$ that are compatible with a given $\mathrm{L}$ according to our rule Eq.(\ref{eq:LfromH}). Indeed, if we choose freely the phases of the off-diagonal entries of $\mathrm{H}$ we can accommodate both time directions, while the diagonal entries can always be taken to be positive by shifting $\mathrm{H}$ by a multiple of the identity. 

Overall, given an undirected unweighted simple graph with $N$ vertices and $E$ edges, $\mathrm{L}$ is completely fixed by the topology and the number of free real parameters of Hamiltonians $\mathrm{H}$ compatible with $\mathrm{L}$ is $N + E - 1$, of which $N-1$ are positive but unbounded real numbers $d_{j}$ for $j=1,...,N-1$, and $E$ are phases $e^{i \phi_{k}}$ with $\phi_{k} \in [ 0, 2 \pi ) $ and $k = 1,...,E $. Despite the fact that all these parameters will contribute non-trivially to the unitary quantum evolution operator $e^{- i \hat{H} t}$, not all of them affect the evolution of transition probabilities between sites. This is best understood by considering a change of basis that sends localized states to localized states, without changing their labels:
\begin{align}
    \vert j \rangle \ \ & \mapsto  \ \ \vert \tilde{j} \rangle := e^{i \alpha_{j}} \vert j \rangle \\
     \langle j \vert \hat{H} \vert k \rangle = \left[ \mathrm{H} \right]_{jk} \ \ &\mapsto \ \ \left[ \tilde{\mathrm{H}} \right]_{\tilde{j} \tilde{k}}  \ = \ e^{i ( \alpha_{k} - {\alpha}_{j})} \left[ \mathrm{H} \right]_{jk}
\end{align}
where $\tilde{\mathrm{H}}$ is the Hamiltonian matrix in the transformed basis. Physical quantities will not be affected: we can profit of this change of basis to cancel some of the phases in $\mathrm{H}$ at the cost of changing relative phases in superposition states, so as to keep all basis-independent quantities unaltered. However, if we restrict ourselves to transition probabilities between sites such as $\pi_{kj}(t) = \vert \langle k \vert e^{- i \hat{H} t} \vert j \rangle \vert^{2}$ and functions thereof, we can neglect the overall phases of the transformed initial and final states. This means that, as far as the $\pi_{kj} (t)$ are concerned, $N-1$ phases of $\mathrm{H}$ will be redundant. Indeed, consider the unitary transformation
$
    U_{g} \left( \underline{\alpha} \right) \ = \ \diag \left( e^{i \alpha_{1} }  , ..., e^{i \alpha_{N}} \right)
$, 
where $\underline{\alpha}$ is an $N$-dimensional real vector that encodes the phases which describe the transformation. The Hamiltonian will change according to $\tilde{ \mathrm{H} } = U_{g}^{\dagger} ( \underline{\alpha} ) \cdot  \mathrm{H} \cdot U_{g}(\underline{\alpha})$, therefore only $N-1$ of the components of $\underline{\alpha}$ actually change $\mathrm{H}$, while an overall phase can be always factored out. Consequently, the number of free parameters in $\mathrm{H}$ that actually affect the evolution of on-site probabilities is $N + (E - N + 1) - 1 = E$, which is just the number of edges. 

To provide a physical intuition for the phase degrees of freedom, we remark that the idea of amending for a local change of phase in the wave-function by changing some parameters in the Hamiltonian reminds of local gauge invariance with gauge group $U(1)$, i.e. of a coupling between a charged particle described by the wavefunction and a classical abelian gauge field. Indeed, we will now argue that the most general unweighted chiral CTQW on a regular lattice and in any dimension may be interpreted as the discretization of the non-relativistic Schr\"{o}dinger equation for a scalar charged particle coupled to a classical electromagnetic field. 
Let us consider a \emph{lattice}, i.e. an infinite graph which tiles periodically a $d$-dimensional Euclidean space. If the graph is regular (each vertex has the same connectivity) then the lattice is called regular too. Since we are  interested in linking quantum walks on lattices to discretized Schr\"{o}dinger equations in a $d$-dimensional space, we require that all the edges have the same weight, reflecting the idea that, apart for the effect of potentials, the transition amplitudes between nearby points in an empty Euclidean space should just depend upon the absolute distance between them. We will refer to a chiral, unweighted QW on a regular lattice as a \emph{homogeneous} (continuous-time) \emph{quantum walk}, and these are our natural candidates to be interpreted as spatial discretizations of quantum theories in Euclidean spaces. 

Let us start by considering the simple scenario of an infinite 2D square lattice. Referring to some arbitrary vertex, we can label each point with a pair of integers $(n, m) \in \mathds{Z}^{2}$ and the corresponding localized state will be $\vert n,m \rangle$. We will denote by $\Psi_{n,m}(t) = \langle n, m \vert \Psi (t) \rangle$ the amplitude of the walker at time $t$ to be found at site $(n,m)$, and by $\Psi (t)$ the vector of the amplitudes over all the sites of the lattice. It is also convenient to explicitly write the rate $\gamma$ that governs the time evolution, so that the Schr\"{o}dinger equation would be symbolically written as ($\hbar = 1$) $
   i  \partial_ t \Psi \ \ = \ \ \gamma \mathrm{H} \cdot \Psi$.
The most general Hamiltonian matrix $\mathrm{H}$ for a CTQW on this lattice will be an infinite, sparse matrix specified by the following entries \footnote{Since the functions $f_{x}(n,m), f_{y}(n,m)$ and $d (n,m)$ are completely unconstrained, this is indeed the most general Hamiltonian with nearest neighbor interactions on a square 2D lattice.}: 
$
   \frac{1}{\gamma} \langle p,q \vert \mathrm{H} \vert n, m \rangle \ = 
   \left[ 4 + d(n,m) \right] \delta_{p,n} \delta_{q,m} 
     - \ \exp[i f_{x} (p-1,q)] \delta_{p,n+1} \delta_{q,m} 
     - \ \exp[-i f_{x} (p,q) ] \delta_{p,n-1}  \delta_{q,m} 
     - \ \exp[i f_{y} (p,q-1)] \delta_{p,n} \delta_{q,m+1} 
     - \ \exp[-i f_{y} (p,q)] \delta_{p,n} \delta_{q,m-1} 
$
where $\delta_{p,n}$ is the Kronecker delta and $f_{x}(n, m), f_{y} (n,m)$ and $d (n,m)$ are real valued functions of the sites positions. 
\par
Now let $a$ be the lattice spacing, which can be assumed to be the same in all directions without loss of generality. By exploring all possible scaling laws of the functions $f_{x}, f_{y}$ and $d$ with $a$, a slight variation of an argument by Feynman (\cite{Feynman}) shows that the only nontrivial continuous limit ($a \to 0$) of this model leads to an Hamiltonian operator $\hat{H}_{c} (\underline{x} )$ of the following form (see Supplementary Material):
\begin{equation}
    \hat{H}_{c} (\underline{x} ) \Psi ( \underline{x} )  \ = \ - K \left( \nabla - i \underline{F} ( \underline{x} ) \right)^{2} \Psi (\underline{x} ) +  U( \underline{x} ) \Psi( \underline{x} )
\end{equation}
where now $\underline{x} \in \reals^{2}$, $U ( \underline{x} ) = \gamma d ( \underline{x} )$, $K =\lim_{a \to 0} a^{2} \gamma$, while $\underline{F} (\underline{x}) = \lim_{a \to 0} \frac{1}{a} \left( f_{x} (\underline{x}), f_{y} (\underline{x} ) \right)$ and $d(\underline{x}), f_{x} (\underline{x}), f_{y} (\underline{x})$ are the continuum generalizations of $d(n,m), f_{x}(n,m)$ and $f_{y}(n,m)$, respectively. The existence of all these limits is a prerequisite to find a nontrivial theory as $a \to 0$. We can now restore $\hbar$, write $U(x) = q V(x)$, $F(x) = q A(x)$ and $K = \frac{\hbar^{2}}{2m}$ in order to arrive at the standard nonrelativistic Schr\"{o}dinger equation for a scalar particle with mass $m$ and charge $q$ in the presence of an electromagnetic field:
\begin{equation}
    i \hbar \ \dfrac{ \partial}{\partial t} \Psi ( \underline{x} ) = - \frac{\hbar^{2}}{2m } \big( \nabla - i q \underline{A} (\underline{x} ) \big)^{2} \Psi( \underline{x} )  + q V( \underline{x} ) \Psi( \underline{x} )
\end{equation}
where $\underline{A} (\underline{x} )$ is the vector potential and $V( \underline{x} )$ is the scalar potential.
{The derivation can be readily generalized to cubic lattices and, even more generally, to regular lattices in any dimension by introducing the appropriate vector potential $\underline{A}(x)$ and with suitable discretizations of the Laplacian \cite{RPB20}, and it shows that chiral CTQWs on regular lattices, despite being the most general of their kind, are in fact always equivalent to discretizations of the Schr\"{o}dinger equation for a scalar particle in the presence of an electromagnetic field, also suggesting a practical implementation for chiral QWs. The reasoning can also be inverted to deduce the form of the phases $f(n,m)$ from a given vector potential. The result is known as Peierls substitution \cite{wer19,yalccinkaya2015two,aidelsburger2015artificial,bernevig2013topological,hofstadter1976energy}, and in our case for a $d$-dimensional cubic lattice it reduces to
$
    f_{j} ( \underline{x} ) \ = \ q \int_{ \gamma } \underline{A} ( \mathbf{r}(t)) dt
$, 
where $\gamma: t \in [ 0, 1] \to \reals^{d}$ is a path from $\underline{x} \in \mathds{Z}^{d}$ to $\underline{x} + a \underline{e}_{j}$ and $\underline{e}_{j}$ is the versor in the $j$-th direction. Notice that this is consistent with $F(\underline{x} )= q A( \underline{x} ) = \lim_{a \to 0} \frac{1}{a} f(\underline{x})$ via the integral mean value theorem.}
\par
{Finally, let us return to gauge invariance. For the 2D square lattice, for each vertex we can cancel the phase of one link attached to it by a phase rotation of the corresponding localized state, therefore we can always set $f_{x}(n,m) = 0$ $\forall n,m \in \mathds{Z}^{2}$, for example. The discretized magnetic field is then given by
$B (n,m) \ = \left[  f_{y} (n+1,m) - f_{y} (n-1,m) \right]/2a$ \footnote{This is a discretized version of $B_{z} = \frac{\partial A_{y} }{\partial x} - \frac{\partial A_{x} }{\partial y}$ for the gauge choice $A_{x} = 0$.},
assuming a symmetric discrete derivative. $B(n,m)$ is to be understood as the component  of the magnetic field at position $(n,m)$, in the \emph{orthogonal direction to the plane}. As an example, assuming a constant $B$ such that $f_{y}(n,m) = n B$, the spectrum of the corresponding $\mathrm{H}$ as a function of the parameter $B$ is the Hofstadter butterfly \cite{hofstadter1976energy}. However, our result is much more general, since it can be stated in any dimension and for any choice of the gauge field and, correspondingly, of the magnetic field.  It also offers a nice interpretation of the \emph{chiral} behavior of chiral CTQWs: in the presence of magnetic fields, the dynamics of a charged particle can be directional and asymmetric under time reversal.}
\par
In conclusion, we have put forward a minimal set of physically motivated 
postulates that a CTQW Hamiltonian should satisfy to properly describe the 
quantum counterpart of a classical RW on a graph. We found that these 
conditions are satisfied by infinitely many quantum Hamiltonians, and 
that any classical RW on a graph corresponds to infinitely many chiral 
CTQWs whose on-site energies are also arbitrary. Our results provide a full characterization of the additional quantum degrees of freedom, available 
to achieve a quantum advantage of CTQWs vs. RWs in specific tasks. 
We also found how to control and manipulate these additional degrees of freedom for a charged walker. The diagonal elements may be determined by the interaction with a classical scalar field, whereas, for regular lattices in generic dimension, the off-diagonal phases may be tuned  with a classical \emph{gauge} field residing on the edges.  
\bibliography{QtoCRW.bib}

\end{document}


\title{Supplementary Material for\\ ``Generalized quantum-classical correspondence for random walks on graphs''}

\author{Massimo Frigerio}
\email{Electronic address: massimo.frigerio@unimi.it}
\affiliation{Quantum Technology Lab $\&$ Applied Quantum Mechanics Group, Dipartimento di Fisica ``Aldo Pontremoli'', Universit\`a degli Studi di Milano, I-20133 Milano, Italy}
\affiliation{INFN, Sezione di Milano, I-20133 Milano, Italy}

\author{Claudia Benedetti}
\email{Electronic address: claudia.benedetti@unimi.it}
\affiliation{Quantum Technology Lab $\&$ Applied Quantum Mechanics Group, Dipartimento di Fisica ``Aldo Pontremoli'', Universit\`a degli Studi di Milano, I-20133 Milano, Italy}
\affiliation{INFN, Sezione di Milano, I-20133 Milano, Italy}

\author{Stefano Olivares}
\email{Electronic address: stefano.olivares@fisica.unimi.it}
\affiliation{Quantum Technology Lab $\&$ Applied Quantum Mechanics Group, Dipartimento di Fisica ``Aldo Pontremoli'',
Universit\`a degli 
Studi di Milano, I-20133 Milano, Italy}
\affiliation{INFN, Sezione di Milano, I-20133 Milano, Italy}

\author{Matteo G. A. Paris}
\email{Electronic address: matteo.paris@fisica.unimi.it}
\affiliation{Quantum Technology Lab $\&$ Applied Quantum Mechanics Group, Dipartimento di Fisica  ``Aldo Pontremoli'',
Universit\`a degli 
Studi di Milano, I-20133 Milano, Italy}
\affiliation{INFN, Sezione di Milano, I-20133 Milano, Italy}

\date{\today}
\maketitle

\section{An intuitive Correspondence via decoherence}
In this Section we aim at providing an operational motivation for the rule:
\begin{equation}
\label{eq:LfromH}
    \left[ \mathrm{L} \right]_{kj} \ \ = \ \ [ \mathrm{H}^{2} ]_{jj} \delta_{jk}  \ - \  \mathrm{H}_{jk} \mathrm{H}_{kj}   
\end{equation}
relating the quantum generator $\mathrm{H}$ and the classical generator $\mathrm{L}$ for 
continuous-time random walks on graphs, and in turn a natural alternative to recover the quantum-classical correspondence via the Born rule. The idea is to consider a temporal coarse-graining, leading to decoherence in the energy eigenbasis. In this way, one prevents the cancellation of the linear term in $t$ in the quantum mechanical evolution by deviating from unitary evolution and introducing decoherence. In particular, let us consider the following Liouville - Von Neumann equation:
\begin{equation}
\label{eqQevDEC}
    \dfrac{ \mathrm{d} \rrho }{ \mathrm{d} t} \ \ \ = \ \ \ - i \left[ \hat{H}, \rrho  \right]  \  - \frac{\gamma}{2} \left[ \hat{H}, \left[ \hat{H}, \rrho \right] \right] 
\end{equation}
describing the so-called intrinsic decoherence of quantum systems \cite{milburn}. To first order in $t$, the probability, according to the Born rule, of finding the walker on site $k$ at time $t$ when starting on site $j$ at time $0$ with the modified quantum evolution (\ref{eqQevDEC}) is:
\begin{align}
     \langle k \vert \rrho (t) \vert k \rangle  =  \langle k \vert j \rangle \langle j \vert k \rangle  +  t \ \dfrac{ \mathrm{d} \langle k \vert \rrho \vert k \rangle}{ \mathrm{d} t } \Bigg\vert_{0}  +  O(t^2)  & =  \ \delta_{jk}   -  \dfrac{ \gamma t}{2} \Big\{ \langle k \vert \hat{H} \big[ \hat{H} , \vert j \rangle \langle j \vert  \big] \vert k \rangle  -  \langle k \vert \big[ \hat{H} , \vert j \rangle \langle j \vert  \big] \hat{H}  \vert k \rangle  \Big\}  \ + \ O(t^2) \   = \notag \\
     & = \ \delta_{jk} \ -  \ \gamma t \left\{ [ \mathrm{H}^{2} ]_{jj} \delta_{jk}  - \vert \mathrm{H}_{jk} \vert^{2}   \right\} \ + \ O(t^2)  
\end{align}
where we used $\rrho_{0} = \vert j \rangle \langle j \vert$ and the Hermiticity of the matrix $\mathrm{H}$ in the last step. Notice that the matrix $\mathrm{L}$ defined by the linear term:
\begin{equation}
\label{eqLaplfromQ}
    \left[ \mathrm{L} \right]_{kj} \ \ = \ \ [ \mathrm{H}^{2} ]_{jj} \delta_{jk}  \ - \  \mathrm{H}_{jk} \mathrm{H}_{kj}   
\end{equation}
is a real, symmetric matrix satisfying: 
\begin{equation*}
 \sum_{k=1}^{n} \mathrm{L}_{kj} \ \ = \ \ [ \mathrm{H}^{2} ]_{jj} \ - \ \sum_{k=1}^{n} \mathrm{H}_{jk} \mathrm{H}_{kj} \ \ = \ \ 0 
\end{equation*}
hence it generates \textcolor{red}{bi-stochastic} transformations. This means that, to first order in $t$, the probabilities computed with the Born rule for any CTQW \emph{with decoherence} are always compatible with some RW.  \\

Notice also that, despite the fact that almost any quantum dynamics can be described by a quantum channel with a semi-group structure (at the level of density matrices), this is not generally true for generic classical stochastic processes. However, we just proved that, whenever the classical stochastic process \emph{has} a semi-group structure, it can be simulated by a quantum system with decoherence in the energy basis. We can state the following:

\begin{propos}
Any classical stochastic process that respects the semi-group structure (i.e., a Markov chain) and has finitely many states can be sampled (for short times) by a quantum computer with decoherence in the energy basis.
\end{propos}

\section{Continuum limit of chiral CTQW on a 2D square lattice}
Let us start from the discrete Hamiltonian for a generic chiral CTQW on an infinite 2D square lattice that was introduced in the main text and we shall also report here for convenience:
\begin{align}
    \frac{1}{\gamma} \langle n', m' \vert \mathrm{H} \vert n, m \rangle \ = \ &  \left[ 4 + d(n,m) \right] \delta_{n',n} \delta_{m',m} \ - \ e^{i f_{x} (n'-1,m')} \delta_{n',n+1} \delta_{m',m} \ - \ e^{-i f_{x} (n',m') } \delta_{n',n-1}  \delta_{m',m} \ + \notag \\
    & - \ e^{i f_{y} (n',m'-1)} \delta_{n',n} \delta_{m',m+1} \ - \ e^{-i f_{y} (n',m')} \delta_{n',n} \delta_{m',m-1} 
\end{align}

Calling $a$ the lattice spacing, we relabel the lattice points as $(n,m) \mapsto (x_{n}, y_{m} )$ where $x_{n} = n a$ and $y_{m} = m a$. By considering $f_{y}$ and $f_{x}$ to be independent functions, we can indeed take the same lattice spacing along $x$ and along $y$, without loss of generality. Now we imagine to extend $\Psi$ to a smooth complex-valued function on the whole $\reals^{2}$ plane, that interpolates between the values of the amplitude at the lattice sites and do the same for the real valued functions $d(x,y)$, $f_{x}(x,y)$ and $f_{y} (x,y)$. With this notation, the action of the Hamiltonian on $\Psi$ could be written as:

\begin{align}
\label{eq:Hxydiscr}
    \frac{1}{\gamma} \left[ \mathrm{H} \cdot \Psi \right] (x,y) \ = \ & \left[ 4 + d(x,y) \right]  \Psi(x,y) \ + \\
    & - \ e^{i f_{x} (x-a,y)}  \Psi(x-a,y) \ - \ e^{-i f_{x} (x,y) } \Psi(x+a,y) \ + \notag \\
    & - \ e^{i f_{y} (x,y-a)} \Psi(x,y-a) \ - \ e^{-i f_{y} (x,y)} \Psi(x,y+a). \notag
\end{align}

First, let us focus on the terms with translations along the $x$ direction, the second line of Eq.(\ref{eq:Hxydiscr}). Since we have to \emph{define} a continuum limit, we can freely postulate the scaling of $f_{x}$ with $a$. If we take $f_{x}$ to be $O(1)$, by expanding the second line up to first order in $a$ we find:
\begin{equation}
    - e^{ i f_{x} } e^{ - i a \partial_{x} f_{x} } \left[ \Psi - a \partial_{x} \Psi \right] - e^{i f_{x}} \left[ \Psi + a \partial_{x} \Psi \right] + O(a^{2} ) \ = \ - 2 \left( \cos f_{x} \right) \Psi + a e^{i f_{x} } \left( \partial_{x} f_{x} \right) \Psi + 2 i a \left( \sin f_{x} \right) \partial_{x} \Psi  + O(a^2)
\end{equation}
where we wrote $f_{x} \equiv f_{x} (x,y)$ and $\Psi \equiv \Psi (x,y)$ for shortness. Taking the limit $a \to 0$ yields a contribution which is diagonal in the position, so it is just a correction to the scalar potential $d(x,y)$. Analogously, if we expanded the last line assuming $f_{y}$ of $O(1)$ in $a$, we would have found a correction term to $d(x,y)$ of the form $- 2 \cos f_{y}$.  Now suppose that $f_{x}$ is $O(a)$, so that we can write $f_{x}(x,y) = a F(x,y)$. Expanding again the second line of Eq.(\ref{eq:Hxydiscr}), this time up to $O(a^2)$, we find:
\begin{align}
    -\big( 1 + i a F_{x} -i a^2 \partial_{x} F_{x} - \frac{1}{2} a^2 F^{2}_{x} \big) \big( \Psi  - a \partial_{x} \Psi + \frac{1}{2} a^{2}  \partial^{2}_{x} \Psi \big)   & -  \big( 1 - i a  F_{x} - \frac{1}{2} a^2 F^{2}_{x} \big) \big( \Psi  + a \partial_{x} \Psi + \frac{1}{2} a^{2}  \partial^{2}_{x} \Psi \big) + O(a^3) \ = \notag \\
    & = \  - 2 \Psi     - a^{2} \Big( \partial_{x} - i F_{x}  \Big)^{2} \Psi  \ + \ O(a^3)  \notag
\end{align}
and $F_{x} \equiv F_{x}(x,y)$. Doing the same for the third line of Eq.(\ref{eq:Hxydiscr}), with $f_{y} = a F_{y}$, the net result is:

\begin{equation}
\label{eq:Hxyexpans}
    \left[ \mathrm{H} \cdot \Psi \right] (x,y) \ = \  - \gamma a^{2} \left( \nabla - i \underline{F}(x,y) \right)^{2} \Psi (x,y)   \ + \ \gamma d(x,y) \Psi(x,y) \ +\ O(a^3) 
\end{equation}
where $\underline{F} = (F_{x}, F_{y})$. If we naively take $a \to 0$ at this point, the result would again be trivial. However, unlike  the previous case, here we can take $\gamma = K/ a^{2}$, with $K \sim O(1)$ and constant, and also $\lim_{a \to 0} \gamma d(x,y) = U(x,y)$, where we are forced to assume $d(x,y) \sim O(a^2)$ to get a finite continuum limit with a potential field landscape. For these choices, the continuum limit is no longer trivial and we arrive at the desired result:
\begin{equation}
\label{eq:Hcont}
     \lim_{a \to 0} \ \  \left[ \mathrm{H} \cdot \Psi \right] (x,y) \ = \  - K  \left( \nabla - i \underline{F}(x,y) \right)^{2} \Psi (x,y)   \ + \ U(x,y) \Psi(x,y)
\end{equation}
We also remark that taking $f_{x} \sim O(1)$ and $f_{y} \sim O(a)$, or vice versa, would prevent $\gamma$ to be $O(a^{-2})$, otherwise the $- 2 \gamma \cos f_{x}$ term would diverge as $a \to 0$. Therefore, a mixed choice of scaling laws for $f_{x}$ and $f_{y}$ yields a diagonal action of $\mathrm{H}$ on $\Psi$, hence a trivial result. The last option is to postulate $f_{x} \sim O(a^{2 + \epsilon})$ for $\epsilon > 0$. The only option to find a nontrivial continuum limit is again for $\gamma = K/a^{2}$ and the result is the same as neglecting $f_{x}$ altogether to begin with. We have thus proved that the only nontrivial continuum limit of a generic chiral CTQW on a square 2D lattice is given by Eq.(\ref{eq:Hcont}). To conclude, let us highlight the fact that the scaling laws $f_{x} \sim O(a)$ and $f_{y} \sim O(a)$ are, in a sense, the most natural choice: these functions define the phases for the hopping amplitudes between neighbors on the lattice, and it is reasonable to require that these phases should go to $0$ as the distance between neighboring points is reduced.